\documentclass[preprint]{myarticle}%

\graphicspath{{figure/}}

\begin{document}
%% ---------------------------
%% title, authors, and abstract
%% ---------------------------

\begin{frontmatter}

% Title
\title{Variability of morphology in beat-to-beat photoplethysmographic waveform quantified with unsupervised wave-shape manifold learning for clinical assessment}

% Authors and affiliations
\author{Yu-Chieh Ho\fnref{label1,label2}}
\author{Te-Sheng Lin\fnref{label1,label2}}
\author{She-Chih Wang\fnref{label3,label4}}
\author{Chen-Shi Chang\fnref{label5}}
\author{Yu-Ting Lin\fnref{label3,label4}}

\fntext[label1]{Mathematics Division, National Center of Theoretical Sciences}
\fntext[label2]{Department of Applied Mathematics, National Yang Ming Chiao Tung University}
\fntext[label3]{Department of Anesthesiology, Taipei Veterans General Hospital}
\fntext[label4]{School of Medicine, National Yang Ming Chiao Tung University}
\fntext[label5]{Department of Anesthesiology, Shin Kong Wu Ho Su Memorial Hospital}

\journal{IOPscience--Physiological Measurement}

% Abstract
\begin{abstract}

We investigated the beat-to-beat fluctuation of the photoplethysmography (PPG) waveform. The motivation is that morphology variability extracted from the arterial blood pressure (ABP) has been found to correlate with baseline condition and short-term surgical outcome of the patients undergoing liver transplant surgery. Numerous interactions of physiological mechanisms regulating the cardiovascular system could underlie the variability of morphology. We used the unsupervised manifold learning algorithm, Dynamic Diffusion Map, to quantify the multivariate waveform morphological variation. Due to the physical principal of light absorption, PPG waveform signals are more susceptible to artifact and are nominally used only for visual inspection of data quality in clinical medical environment. But on the other hand, the noninvasive, easy-to-use nature of PPG grants a wider range of biomedical application, which inspired us to investigate the variability of morphology information from PPG waveform signal. We developed data analysis techniques to improve the performance and validated with the real-life clinical database.

\end{abstract}

\end{frontmatter}

%% ---------------------------
%% sections
%% ---------------------------

\begin{abstract}

We investigated the beat-to-beat fluctuation of the photoplethysmography (PPG) waveform. The motivation is that morphology variability extracted from the arterial blood pressure (ABP) has been found to correlate with baseline condition and short-term surgical outcome of the patients undergoing liver transplant surgery. Numerous interactions of physiological mechanisms regulating the cardiovascular system could underlie the variability of morphology. We used the unsupervised manifold learning algorithm, Dynamic Diffusion Map, to quantify the multivariate waveform morphological variation. Due to the physical principal of light absorption, PPG waveform signals are more susceptible to artifact and are nominally used only for visual inspection of data quality in clinical medical environment. But on the other hand, the noninvasive, easy-to-use nature of PPG grants a wider range of biomedical application, which inspired us to investigate the variability of morphology information from PPG waveform signal. We developed data analysis techniques to improve the performance and validated with the real-life clinical database.

\end{abstract}
\section{Introduction}

The human cardiovascular system fluctuates with time, so is the signal waveform. Researches showed the morphology in the arterial blood pressure (ABP) waveform is both dynamical and sophisticated. Extracting its information has facilitated various applications in the clinical medicine \cite{teboul2016less, avolio2009role, hatib2018machine}. Recently we have reported a new kind of the ABP waveform information, the beat-to-beat \textit{variability of morphology} (var. of morph.), which is associated with the condition of the patient undergoing liver transplant surgery and their short-term outcome\cite{wang2023arterial}. 

From the tens of thousands of pulses in a continuous 10-hour ABP signal waveform of the same surgical patient, we have observed that no two pulses could be considered identical in terms of waveform morphology\cite{lin2021intraoperative, wang2020novel, shen2022robust}. While the morphology of an ABP waveform cycle is determined by the wave reflection of the blood flow from the heart to the whole vascular tree in the human body\cite{vlachopoulos2011mcdonald}, the variability of morphology could reflect the numerous interactions between various physiological mechanisms everchanging to regulate the cardiovascular system \cite{wang2023arterial}. 
The above finding and the literature on the ABP waveform \cite{wang2023arterial, shen2022robust,lin2021wave} inspire us to investigation on the waveform of the photoplethysmography (PPG) in the present study. 

Although both ABP waveform and PPG waveform signals display pulsating waveform signals, they are different. The invasive intra-arterial blood pressure measure allows direct pressure measurement as well as waveform information in absolute unit via the connecting pipe principle. Hence, ABP waveform information has be used to assess various hidden conditions of the cardiovascular system \cite{sluyter2019identification, wang2020novel}. On the other hand, the non-invasive PPG relies on the relative difference in different wavelength light absorption, which requires frequent automatic adjustment in the signal processing stage to obtain the arterial oximeter and the pulsation waveform displayed on the monitor. Therefore, while the oximeter readings is indispensable in various situation of the clinical medicine, its waveform signal is more susceptible to the interference from various external factors and generally considered to be less reliable\cite{lin2019unexpected}.

The Dynamic Diffusion Map (DDMap) algorithm was developed to tackle the multivariate nature of the cardiovascular waveform morphology \cite{wang2020novel, lin2021wave, shen2022robust}. By treating each one segment of the waveform within a heart-beat cycle as a data point in high dimensional space, DDMap yields the hidden structure of the data, which subsequently grants the observation and quantification of the variability of morphology information. It is worth mentioning that the DDMap algorithm possesses the abilities of revealing the inner structure as well as the abilities of robustness in statistics \cite{shen2022robust, shen2022scalability}. % Analyzing the ABP waveform data during the surgery via this technique, we have reported its variability of morphology is associated with the clinical condition of the liver transplant recipient, even the surgical outcome. 

\begin{figure}[!t]
  \centering
  \includegraphics[width=\textwidth]{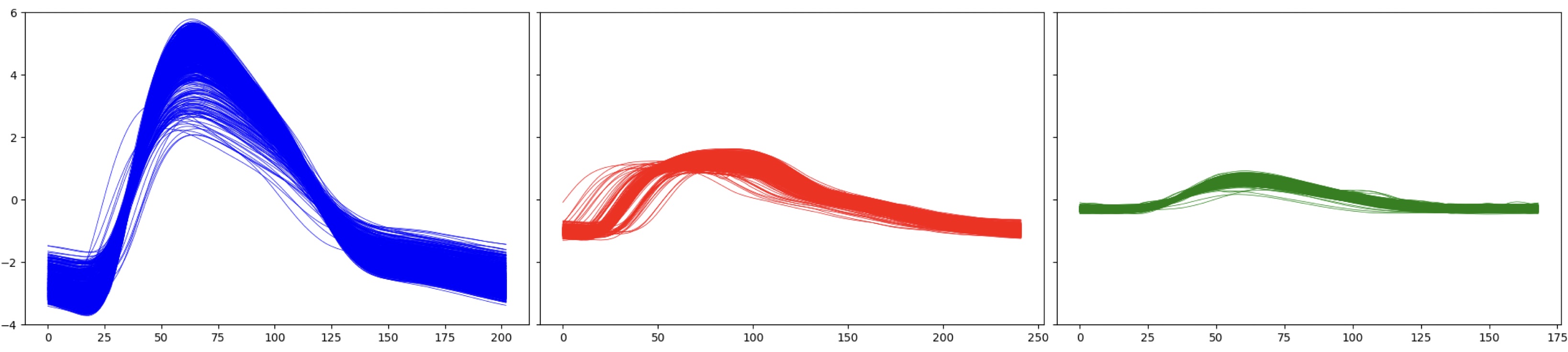}
  \caption{The original beat-to-beat pulses of three cases. Note that they share the same y-axis range.}
  \label{fig:artifact}
\end{figure}

We hypothesized that the quantitative variability of morphology information could be obtained from this physical modality. The motivation is that the non-invasive and ubiquitous PPG could grant applications to wider biomedical situations than the direct intra-arterial blood pressure waveform recorded exclusive in the operating room or the critical care unit in the hospital. However, we still need to deal with several PPG signal artifact problems, see Figure \ref{fig:artifact}.

The original beat-to-beat pulses between three different cases in Figure \ref{fig:artifact} is used to reveal the PPG signal artifacts such as: (1) Different cases and pulses have different baselines; (2) The pulse dynamics are not significant and are divergented between each cases and pulses; (3) The unavoidable noise interference of pulses while recording; and (4) The dispersion of pulses of a case cause by dynamical physiological conditions. The first two problems can be eliminated by our standard preprocessing procedure. As for the third and fourth problems, we propose the combination of windowing method and the Wasserstein-1 distance with the DDMap algorithm to attenuate these artifacts. The effectiveness of the variability of morphology information in the PPG signal is demonstrated with clinical data.

\section{Methodology}

In this section, we elaborate the standard quantification of the variability of morphology and the techniques we used to deal with the PPG signal artifact. In Section \ref{subsec:2-1}, the collected PPG waveform is preprocessed in order to adopt the unsupervised manifold learning technique, DDMap, which is fully explained in Section \ref{subsec:2-2}. Section \ref{subsec:2-3} clarifies how we quantify the waveform morphology. Section \ref{subsec:2-4} describes the standard procedure of obtaining the quantitative variability of morphology. Section \ref{subsec:2-5} presents techniques that aim to decrease the influence of signal artifacts and obtain better performances. Finally, statistical analysis and sensitivity analysis between the variability of morphology and the clinical scores systems is stated in Section \ref{subsec:2-6} and Section \ref{subsec:2-7}, respectively.

\subsection{Preprocessing of PPG waveform}
\label{subsec:2-1}

The continuous physiological waveform dataset was collected from a single center prospective observational study between 2018 and 2021 in Taipei Veterans General Hospital, Taipei, Taiwan. 85 living donor liver recipients were recruited after Institutional Review Board approval (IRB No.: 2017-12-003CC and 2020-08-005A) and written informed consent obtained from each patient. 
The four signals of these 85 cases, including ABP, PPG, central venous pressure (CVP), and electrocardiogram (ECG), were collected from the patient monitor (GE CARESCAPE\textsuperscript{TM} B850, GE Healthcare, Chicago, IL) via the data collection software S5 Collect (GE Healthcare). In this study, we focus on the PPG signal to see how it performs with respect to the ABP signal. 

To obtain consecutive pulses from continuous PPG waveform at 300 Hz sampling rate, each pulse is automatically identified with the maximum of the first difference of the systolic phase, which is the ascending part of the pulse. The whole pulse waveform could be isolated accordingly in most situation. Next, pulses that have maximum or minimum values not within reasonable range, or those that contain long straight lines, signifying the signals have not been detected for a long period of time, are regarded as poor quality pulses.
These pulses are automatically identified, removed, and replaced by using linear interpolation with respect to their time locations. Lastly, each pulse is subtracted by its median as the baseline, then divided by its $\ell^2$-norm in order to increase its variability of morphology and decrease the difference between pulses made by artifacts when recording the waveform. Furthermore, they are truncated to the same length of 140 for the sake of the subsequent works. 

After the basic preprocessing, we remove the signal artifacts of the non-aligned pulses' baseline between different cases, and the pulses dynamics' shortage and divergence. 
But in some rare occasion, as we see in the middle of Figure \ref{fig:example} and the left figure of Figure\ref{fig:improve}(a), the pulse waveform could also be modulated due to the inevitable transient noise in the signal, or possibly a dynamical physiological condition rendered an atypical shape of the ascending part of the pulse. We encounter this situation more frequently than in the ABP waveform analysis in our previous work~\cite{wang2023arterial}, which inevitably interfere the identification of the systolic phase mentioned above. That is the technical issue we would addressed later.

\subsection{Unsupervised manifold learning technique}
\label{subsec:2-2}

The beat-to-beat variation of the morphology pulses within each heart beat cycle is too subtle and sophisticated to be observed with the naked eye. Accordingly, we treated each pulse as a high-dimensional data point, and the DDMap\cite{lin2021wave} is utilized to find a lower-dimensional representation of the point cloud to visualizes the relationships between beat-to-beat pulses in the high-dimensional space. 
The pseudo-code of the DDMap algorithm is presented in Algorithm \ref{algo:DDMap}. In the algorithm, there are only one manually chosen parameters $0<q\ll 140$, which determines the dimension of the DDMap embedding. We empirically set $q=15$ whenever we use the DDMap algorithm. Note that our input dataset $X$ is a sequence of beat-to-beat pulses, and thus the output embedding $\Psi$ informs the time sequence of the pulses. This allow us to analyze the waveform dynamics using the trajectory of DDMap embedding that evolve through time.

\begin{algorithm}[!t]
\caption{The pseudo-code of the Dynamic Diffusion Map (DDMap).}
\label{algo:DDMap}
\begin{algorithmic}[1]
\Require $X=\{x_i\}_{i=1}^n \subset \mathbb{R}^{140},\ 0<q\ll 140$.
\Ensure $\Psi = \{\Psi_i\}_{i=1}^n \subset \mathbb{R}^q$.

\State Construct an affinity matrix $\displaystyle W_{ij} = \exp(-\frac{\|x_i-x_j\|_{\ell^2}^2}{\varepsilon})\quad x_i\in X$, and $\varepsilon$ is the 25-th percentile of all pairwise points in $X$.

\State Construct a diagonal matrix $D$ where $D_{ii}$ is the $i$-th row sum of $W$.

\State Compute the SVD of $D^{-1}W=U\Lambda V^T$. Preserve only the $(q+1)$-largest eigenpairs, then discard the largest one.

\State Construct the DDMap embedding $\Psi_i:x_i\to e_i^TU\Lambda$ for $i=1,2,...,n$.

\end{algorithmic}
\end{algorithm}

The DDMap algorithm works as following. In step 1, we construct a weighted graph in form of the affinity matrix that forms on the high-dimensional dataset, where an edge is close to 1 if the end points of the edge are close to each other in the high-dimensional $\mathbb{R}^{140}$ space, and is close to 0 otherwise. And in step 2, a diagonal matrix with row sum of the affinity matrix. In step 3, we first explicate the matrix $D^{-1}W$ then clarify why we use singular value decomposition on it. The matrix $D^{-1}W$ is a transition matrix (Markov chain), since all of its entries are nonnegative and all of its rows sum to 1. Note that the $ij$-th element of a transition matrix represents the probability of transition in one time step from $x_i$ to $x_j$ on the graph, $p(x_i,x_j)$. Now, we obtain the {\it diffusion distance (DDist)} pertaining to DDMap, as $DDist(x_i,x_j):=\|\sum_{x_k\in X}p(x_i,x_k)-p(x_k,x_i)\|$. $DDist(x_i,x_j)$ is small when there is a large number of short paths on the graph that connect $x_i$ and $x_j$, and vice versa. The diffusion distance can be directly linked to the spectral properties as $DDist(x_i,x_j):=\|\sum_{x_k\in X}p(x_i,x_k)-p(x_k,x_i)\| \approx \| e_i^TU\Lambda - e_j^TU\Lambda \|.$ Note that the eigenvalues of $D^{-1}W$ is $1=\lambda_1>\lambda_2\geq\lambda_3\cdots\geq\lambda_n\geq -1$, the eigenvector corresponding to eigenvalue 1 is a constant vector, and an eigenvalue reflects the importance of its corresponding eigenvector. Therefore, we may discard the first eigenvector and choose a proper $q$ to embed the dataset into a much lower-dimensional euclidean space. As we construct the embedding as in step 4, we are actually obtaining an embedding such that the euclidean distance between pairwise points in the low-dimensional space is roughly equal to the $DDist$ between those points.

\begin{figure}[!t]
  \centering
  \includegraphics[width=\textwidth]{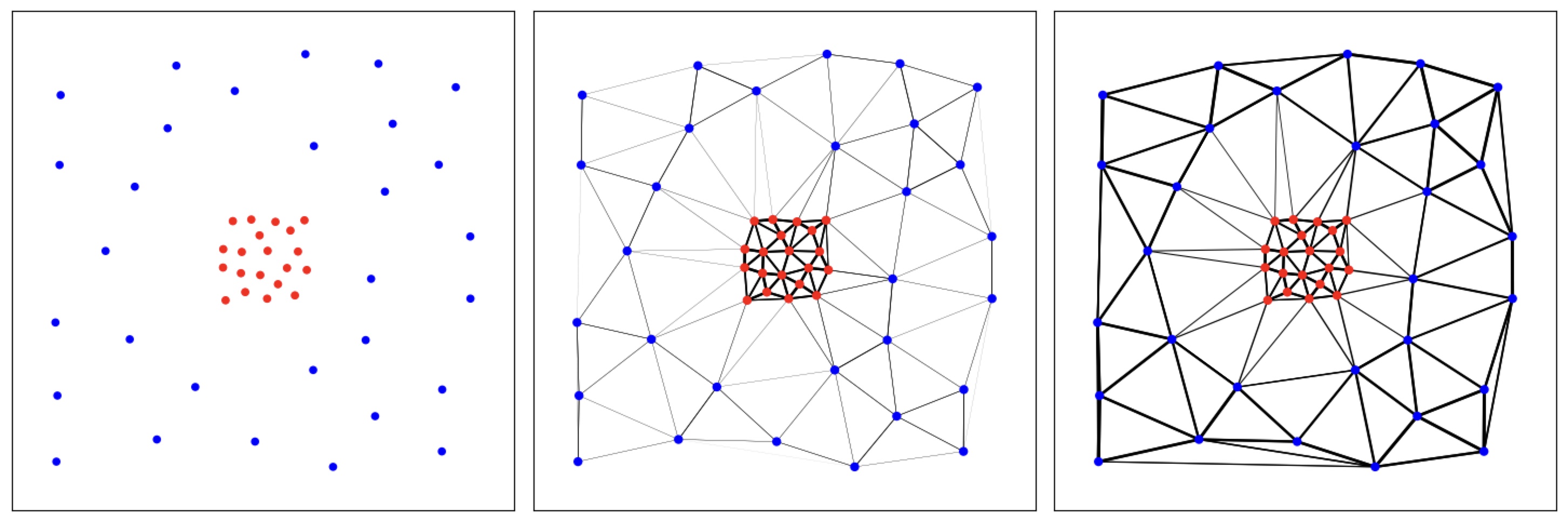}
  \caption{Example of using local scaling in constructing affinity matrix in the DDMap algorithm. Left: A dataset with two groups of points (red and blue) having different densities. Middle: Affinities between data points using global scaling. The black line between pairwise points represent the affinity between them. The thicker it is, the stronger the affinity between those pairwise points are, vice versa. Right: Affinities between data points using local scaling. In this figure, we can see that the affinity between the blue points is stronger with local scaling compared to global scaling, and that the affinity between red points and blue points is relatively weaker. In this case, we can easily separate two groups of points using DDMap algorithm since both groups are strongly connected inside the groups and are poorly connected between two groups.}
  \label{fig:local-scaling}
\end{figure}

In the case when a dataset contains clusters with different densities, which most of the real-world dataset does, an affinity matrix using a global bandwidth in the DDMap algorithm may fail to present the real connectivity between points. To avoid this, local-scaling bandwidths \cite{zelnik2004self} is used to construct a better affinity matrix instead. That is, we change $\varepsilon$ in step 2 of Algorithm \ref{algo:DDMap} to $\|x_i-x_{s}\|_{\ell^2}$, where $x_s\in X$ is the $s$-th nearest neighbor of $x_i$. Following the suggestion in \cite{zelnik2004self} and taking into account the size of our dataset, we choose $s=15$ when using Algorithm \ref{algo:DDMap} with local-scaling. See Figure \ref{fig:local-scaling} for the difference between using a global bandwidth and local-scaling bandwidths in an affinity matrix. 
From here on, when we mention the DDMap algorithm or the DDMap embedding, we means the algorithm \ref{algo:DDMap} with local scaling and its' result.

\subsection{Calculation of the variability of morphology}
\label{subsec:2-3}

\begin{figure}[!p]
  \centering
  \includegraphics[width=\textwidth]{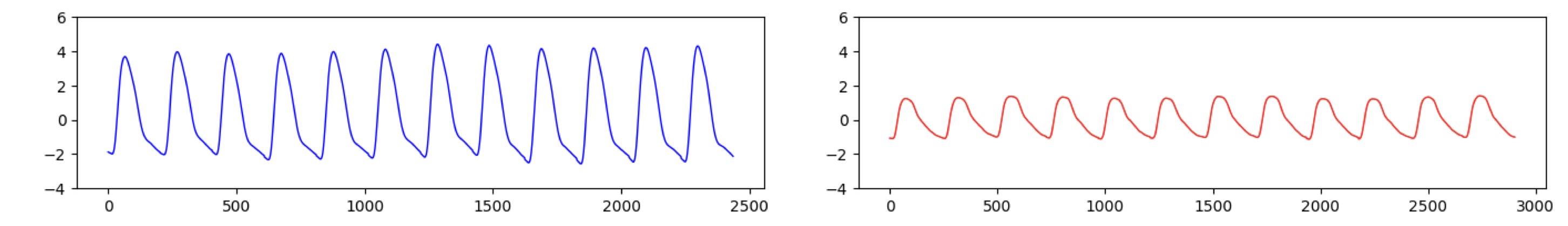}
  \\$ $
  \includegraphics[width=\textwidth]{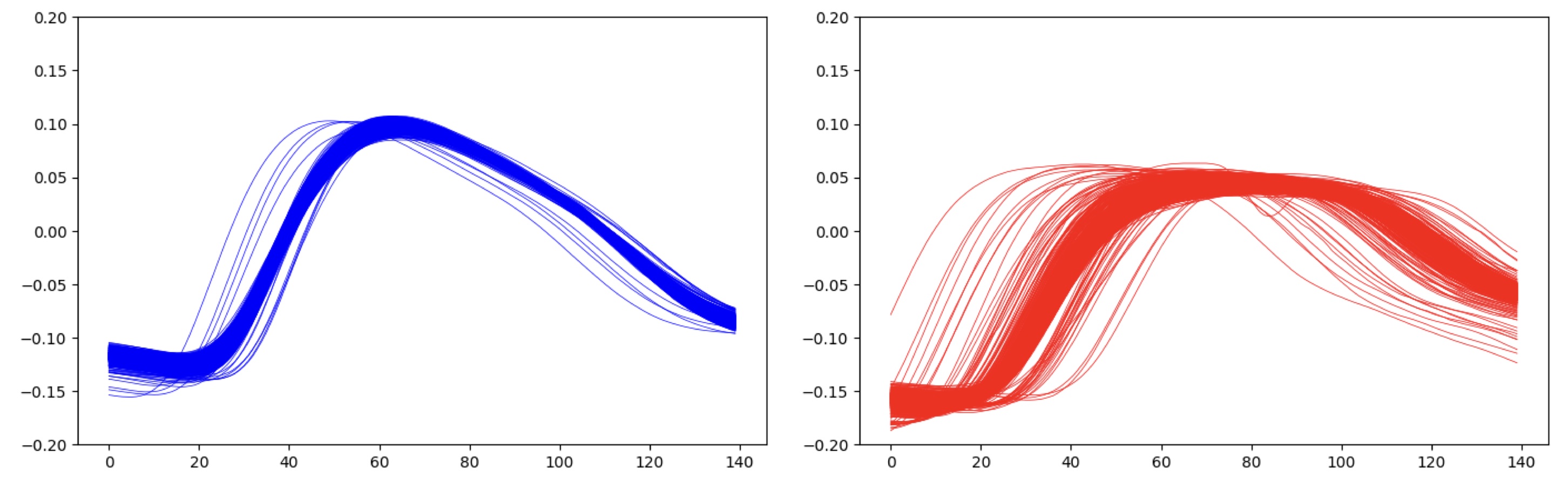}
  \\$ $
  \includegraphics[width=\textwidth]{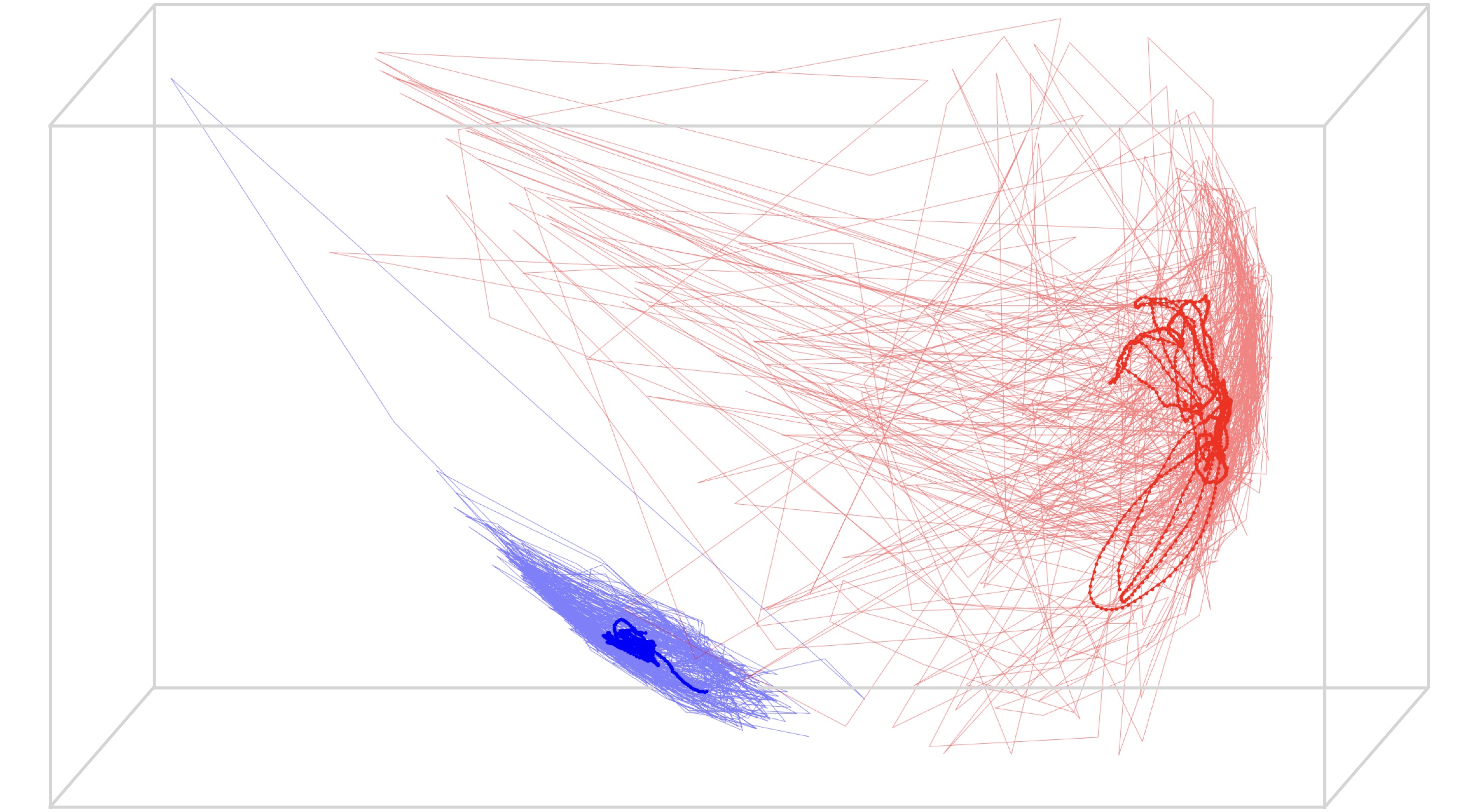}
  \caption{From top to bottom: Continuous PPG waveform before preprocessing, consecutive pulses after preprocessing, and DDMap embedding of two cases. In the embedding, the light color paths indicate the trajectories, and the dark color paths indicate the trends. The blue case has Meld\_Na score 56, and the red case has Meld\_Na score 5. In the figure of embedding, we can see that a case with high Meld\_Na score does not manifest intense dynamical evolution of pulses, which also gives a slow evolution of trend points. On the other hand, a case with low Meld\_Na score gives a lot more complex dynamical change, and thus a fast evolution of trend points. We may quantify the dynamical evolution of trend points of cases then compute their correlations between Meld\_Na score, etc. The results of PPG signals expect to be similar to the results of ABP signal.}
  \label{fig:example}
\end{figure}

The DDMap embedding and its trajectory provides a concise overview of the complex dynamical evolution. We further apply a moving median followed by a moving mean filter to obtain the trend of trajectory. Suppose a case has $L$ pulses and its embedding points are $\{\Phi_i\}_{i=1}^L$, then the trend $T$ of the embedding is 
\begin{equation}
    T_i = \frac{1}{k}\sum_{m=i-k+1}^{i}\text{median}(\Phi_{m-(k+1)/2},...,\Phi_{m+(k+1)/2})\quad \text{for }i=1,...,L,
\end{equation}
where $k$ is chosen manually. Note that those $\Phi_i$ with $i<1$ or $i>L$ were removed from the median pool, same as for $m$. The fraction in front of the summation depends on the number of $m$ that are summing up. See Figure \ref{fig:example} for the visualization of two cases' original signals without preprocessing, beat-to-beat pulses after preprocessing, their DDMap embedding, and their trends. 

Now, by exploiting the DDMap embedding trajectory and its trend, we may quantify the variability of morphology. For each case, the variability of morphology 
\begin{equation}
    \text{var. of morph.} :=\frac{1}{L-1}\sum_{i=2}^L \|T_i-T_{i-1}\|,
\end{equation}
which captured the slow-vary evolution, is calculated as the mean of distances between each consecutive trend points, quantifying how fast the trend evolves. The variability of morphology measures are intuitively derived from the trajectory structure to assess the amount of waveform dynamics. 

The trend preserved the relative slow movement component that is more relevant to the inner dynamics of the cardiovascular system according to our previous study\cite{wang2023arterial}. As the fast movement part is often elicited by the variation of the venous blood returning to the heart due to respiratory cycle, arrhythmia such as premature contracture, atrial fibrillation, or even the transient motion artifact at the signal acquisition stage, the physiological regulation mechanisms exert controls on the cardiovascular system at the time scale longer than the breathing cycle\cite{vlachopoulos2011mcdonald}.

% \begin{figure}[!t]
%   \centering
%   \includegraphics[width=\textwidth]{figure/2-2.jpg}
%   \caption{Partial trajectory and trend of Figure \ref{fig:example} red case. The arrow indicates the time direction. The "pulse-wise" variability of morphology are shown as the blue lines.}
%   \label{fig:trend,TM,FM}
% \end{figure}

\subsection{Standard procedure of obtaining variability of morphology}
\label{subsec:2-4}

First of all, we preprocess the PPG waveform into consecutive pulses as we described in Section \ref{subsec:2-1}. One case is removed due to the shortness of pulse length after trimming, and two cases are removed since they do not have enough legitimate pulses. Accordingly, there are 82 cases of neohepatic phase. 

To let the quantitative indices comparable and convenient in future applications, we use presurgical PPG data of 85 cases as reference baseline dataset as we have done previously\cite{wang2023arterial}.  

For pulses of neohepatic phase of each case, we consider them as a non-baseline dataset and compute them individually. A dataset that combines the baseline dataset and a non-baseline dataset is formed, and a DDMap embedding that contains embedding of baseline dataset and non-baseline dataset is obtained by running the DDMap algorithm. Then, we compute the trends and variabilities of morphology of baseline dataset and non-baseline dataset. To rescale the final non-baseline dataset's variability of morphology, we consider the formula 
\begin{equation}
    \text{var.\ of\ morph.}^\ast = \frac{\text{var. of morph.} -\text{median}(\text{pool of var. of morph.})}{\text{IQR}(\text{pool of var. of morph.})}\times25+60, 
\end{equation}
where pool of var. of morph. contains all variabilities of morphology of the baseline dataset, and IQR is the abbreviation for interquartile range.

\subsection{Techniques to deal with PPG signal artifact}
\label{subsec:2-5}

Regarding noise effects and beat-to-beat pulses dispersion that cannot be eliminated by standard perprocessing procedures, we additionally consider two techniques to deal with them. The first technique is to replace the Euclidean distance $d(x_i,x_j) = \|x_i-x_j\|_{\ell^2}$ with the Wasserstein-1 distance $d_{w_1}(x_i,x_j)$ \cite{vaserstein1969markov} in step 2 of Algorithm \ref{algo:DDMap}, which is out of consideration for removing the dispersion characteristic. The second technique is the Hamming window \cite{smith2011spectral}, which deal with both noise and dispersion artifact, and is further applied in the signal preprocessing step. 

\begin{figure}[!t]
  \centering
  \includegraphics[width=\textwidth]{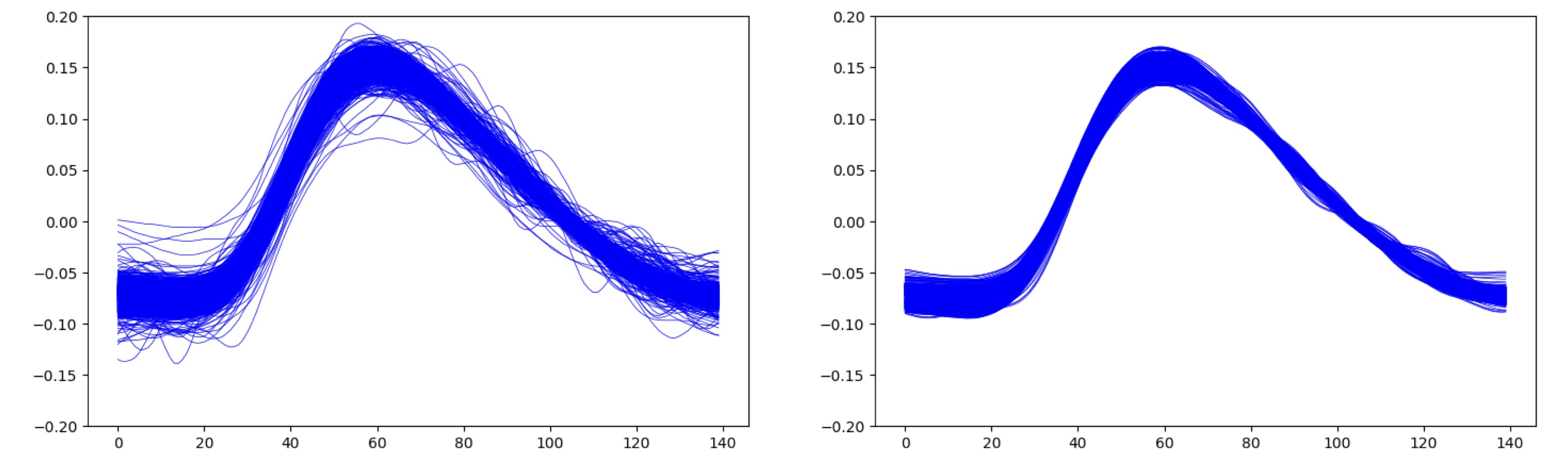}
  \newline (a)
  \includegraphics[width=\textwidth]{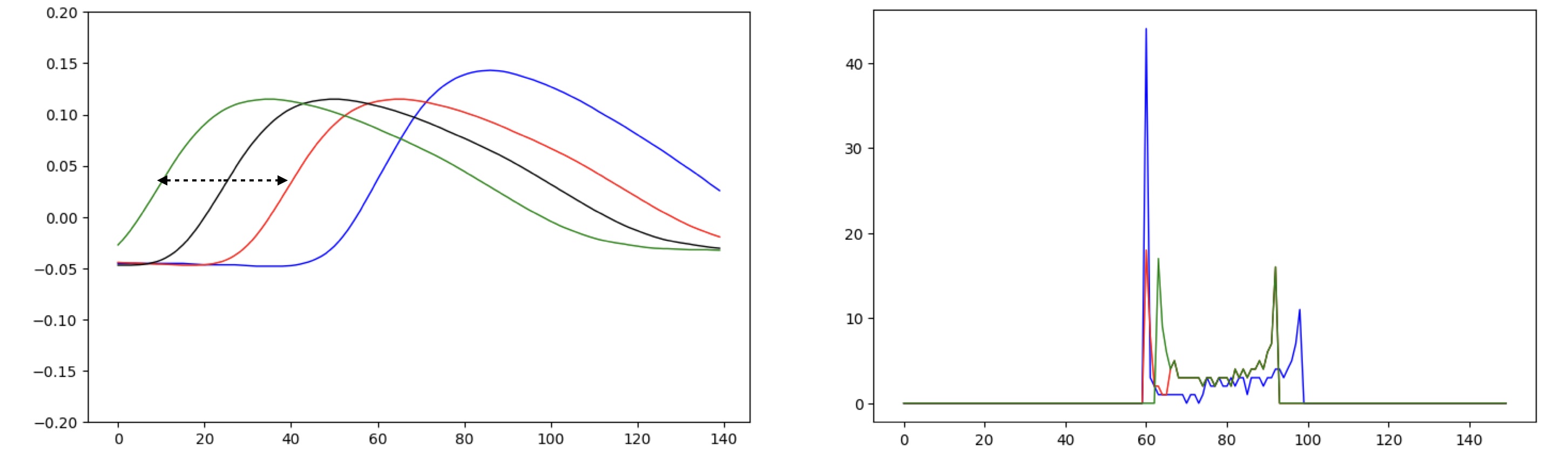}
  \newline (b)
  \caption{(a) Example of using window technique. Left: Pulses without windowing. Right: Pulses with Hamming window of $w=11$. This technique reduce the noise and dispersion of a signal by not only consider the signal itself but also it's nearby signals. (b) Example of using Wasserstein-1 distance. In the left figure, the red and green lines are non-align pulses, which is originally located at the black line, and the blue line is another pulse. Ideally, the distance between red and green lines in high-dimensional space should be closer than the distances between both lines and blue line. If we use Euclidean distances to compute the affinity matrix, then the distance between red and blue lines is $0.68$, and the distance between green and blue lines is $1.32$. However, the distance between red and green lines is $0.86$. On the other hand, if we use Wasserstein-1 distance, the distance between red and blue lines is $34.50$, the distance between green and blue line is $52.74$, and the distance between red and green lines is $26.57$. Accordingly, the Wasserstein-1 distance between non-align pulses is smaller than the Wasserstein-1 distances between both non-align pulses to the blue pulse. The right figure shows the level sets of the three pulses. We can see the level set of red and green pulses are more similar than the level set of blue pulse. In fact, red and green lines coincide after $x=66$.}
  \label{fig:improve}
\end{figure}

The Hamming window is defined as $W_i=0.54-0.46\cos(2\pi(\frac{i}{w-1}+\frac{1}{2}))$, where $-\frac{w-1}{2}\leq i\leq \frac{w-1}{2},\ w$ is an odd number. Suppose the neohepatic phase of a case has pulses $\{x_i\}_{i=1}^L$, then applying the Hamming window to this case means to replace each pulse $x_i$ by $\frac{1}{w} \sum_{j=-(w-1)/2}^{(w-1)/2} W_jx_{i+j}$. Note that those $x_{i+j}$ with $i+j<1$ or $i+j>L$ were removed from the summation. The fraction in front of the summation depends on the number of $j$ that are summing up. See Figure \ref{fig:improve}(a) for the effect of the Hamming window on pulses of a phase. This technique considers no only the signal itself, but also its nearby signals, thereby reduces the signal noise and dispersion. 

For the purpose of computing the Wasserstein-1 distance between pulses, we define a level set with lines parallel to the $x$-axis, $y$-range be set to $[-0.35,0.4]$, and the step of each level is $0.05$. For pulses $x_i$ and $x_j$, we compute the cumulative distribution functions $F_{x_i}$ and $F_{x_j}$ of their level sets, then the Wasserstein-1 distance $d_{w_1}(x_i,x_j)=\sum_{t=1}^{150} |F_{x_i}(t)-F_{x_j}(t)|$. 
See Figure \ref{fig:improve}(b) for an example of why using the Wasserstein-1 distance in the Algorithm \ref{algo:DDMap} improves performance.

\subsection{Statistical analysis}
\label{subsec:2-6}
After obtaining the variability of morphology of the neohepatic phase of all cases, the Spearman correlation coefficient (CC) is used to measure the linear relationship between the variabilities of morphology and the clinical scores. 
Since the underlying distribution of the indices is unknown, the bias-corrected and accelerated bootstrap using $100,000$ random samplings with replacement is exploited to establish the $95\%$ confidential interval of each CC. Also, a test of a null hypothesis that the distributions underlying the samples are uncorrelated and normally distributed is performed, and the $p$-value is reported.

To investigate the effect of the Hamming window technique, we run the procedure on the original dataset, which is obtained directly after the preprocessing, and on another dataset where the Hamming window is applied. We examine the difference between results of Algorithm~\ref{algo:DDMap} with affinity matrix using Euclidean distance and Wasserstein-1 distance. Also, we add the results of ABP signal obtained from the previous research to compare the similarity of presentations among ABP and PPG signals. Accordingly, there will be performances of five different models to discuss in this section. 

For convenient, we named the five models as follows: model 1 is the case where the original dataset and the Euclidean distance are used; model 2 is the case where the hamming window dataset and the Euclidean distance are used; model 3 is the case where the original dataset and the Wasserstein-1 distance are used; model 4 is the case where the hamming window dataset and the Wasserstein-1 distance are used; and model ABP, which is shown for comparing the results between ABP and PPG signals.

We compare the variability of morphology of five models with the revised Model for End-Stage Liver Disease (MELD\_Na) \cite{brown2002model,biggins2006evidence} and the early allograft failure (EAF) scores, including L-GrAFT10 \cite{agopian2021multicenter,agopian2018evaluation} and the EASE score \cite{avolio2020development}. These score systems has been developed by the combination of laboratory examination results to access the liver disease acuity of a patient. The higher the MELD\_Na score means higher priority for liver transplant surgery. Similarly, higher L-GrAFT10 and the EASE scores suggest worsen outcome after the transplant surgery. As higher variability of morphology is associated with better condition, which means lower clinical scores, the theoretical perfect CC is $-1$.

\subsection{Sensitivity analysis}
\label{subsec:2-7}

In the whole procedure, there are two manually chosen parameters: Hamming window size $w$ whenever we uses the Hamming window technique, and trend step size $k$ when calculating trends. Sensitivity analysis is created to exam how variations in the uncertain parameters $w$ and $k$ affect the performances of the procedure, and for testing the robustness of the performance in the presence of uncertainties. 
Note that when there are two input uncertainties, it involves calculating how much the performance of procedure changes when we make an adjustment to one of its input variables while keeping another as a constant. 
\section{Result}

\subsection{Statistical analysis}

\begin{figure}[!t]
  \centering
  \includegraphics[width=\textwidth]{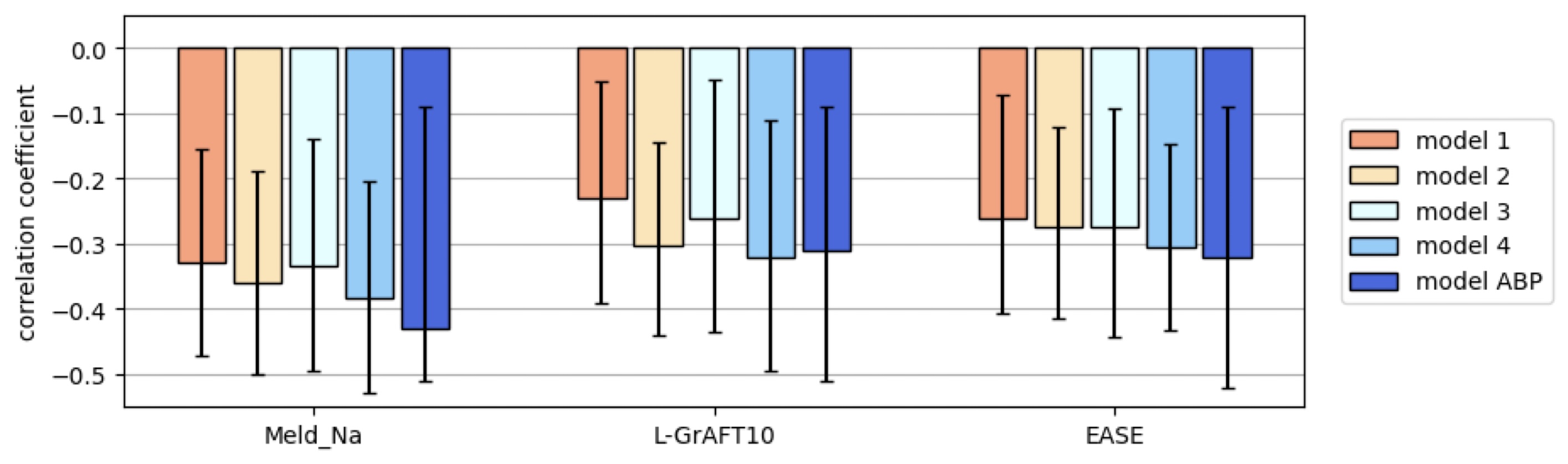}
  \caption{The visualization of the Spearman correlation coefficients and the 95$\%$ confidential intervals between variability of morphology of five models and three clinical scores.}
  \label{fig:result}
\end{figure}

The visualization of the CCs and the $95\%$ confidential intervals between variability of morphology of five models and clinical scores, including MELD\_Na, L-GrAFT10, and EASE score, are presented in Figure \ref{fig:result}. The detail of the results, which we mentioned in the Section \ref{subsec:2-6}, will shown in the form $(a1, [a2, a3], a4, a5)$ with explicit values, where $a1$ is the correlation coefficient, $[a2, a3]$ is the 95\% confidential interval, $a4$ is the $p$-value of the null hypothesis test, and $a5$ is the number of comparing cases. 
The null hypothesis test is consider notable(mark as $^\ast$) when the $p$-value is less than $0.01$, and is consider significant(mark as $^{\ast\ast}$) when the $p$-value is less than $0.001$.

The results between MELD\_Na score and the variability of morphology of the five models are: 
model 1 (-0.329, [-0.47, -0.16], 0.003$^{\ast}$, 82); 
model 2: (-0.359, [-0.50, -0.19], 0.001$^{\ast}$, 82); 
model 3: (-0.333, [-0.49, -0.14], 0.002$^{\ast}$, 82); 
model 4: (-0.384, [-0.53, -0.20], 0.00037$^{\ast\ast}$, 82); and 
model ABP: (-0.430, [-0.62, -0.21], 0.00005$^{\ast\ast}$, 83). For MELD\_Na score, $p$-values are all notable or significant. The CCs and $p$-value both indicates that the correlations between the five models' variability of morphology and the MELD\_Na score are strong.

The results between L-GrAFT10 score and the variability of morphology of the five models are: 
model 1: (-0.230, [-0.39, -0.05], 0.038, 82); 
model 2: (-0.303, [-0.44, -0.14], 0.006$^{\ast}$, 82); 
model 3: (-0.261, [-0.44, -0.05], 0.018, 82); 
model 4: (-0.321, [-0.49, -0.11], 0.003$^{\ast}$, 82); and 
model ABP: (-0.310, [-0.51, -0.09], 0.003$^{\ast}$, 83). 

The results between EASE score and the variability of morphology of the five models are: 
model 1: (-0.261, [-0.41, -0.07], 0.019, 80); 
model 2: (-0.274, [-0.41, -0.12], 0.014, 80); 
model 3: (-0.275, [-0.44, -0.09], 0.014, 80); 
model 4: (-0.306, [-0.42, -0.15], 0.006$^{\ast}$, 80); and 
model ABP: (-0.320, [-0.52, -0.09], 0.0033$^{\ast}$, 81). 

For the four PPG models we performed, using either the technique of Hamming window (model 2) or Wasserstein-1 distance in the DDMap algorithm (model 3) gives better results compare to the result of standard procedure (model 1). Most of the time, using only Hamming window is more effective than using only Wasserstein-1 distance. Thus, it is normal to think that using both Hamming window and Wasserstein-1 distance (model 4) gives the best performance, and the results do confirm this conclusion. That is, the performance of model 4 gives best CCs between PPG variability of morphology and all three clinical scores. The CCs all exceed -0.3, and the $p$-values are all notable or significant. 

We compare the results of best PPG model (model 4) with the results of model ABP. The CCs between the model 4's variability of morphology and the model ABP's variability of morphology are all exceptionally similar above our expectation, since their CC differences are only up to a gap of $\pm 0.05$. Note the in the case of L-GrAFT10 score, CC of model 4 even exceed CC of model ABP. Also, the 95\% confidential intervals of model 4 are all shorter with respect to model ABP, which indicates a more precise CC of model 4.

\subsection{Sensitivity analysis}

\begin{figure}[!p]
  \centering
  \begin{tabular}{cc}
  (a) & \raisebox{-.5\height}{\includegraphics[width=.95\textwidth]{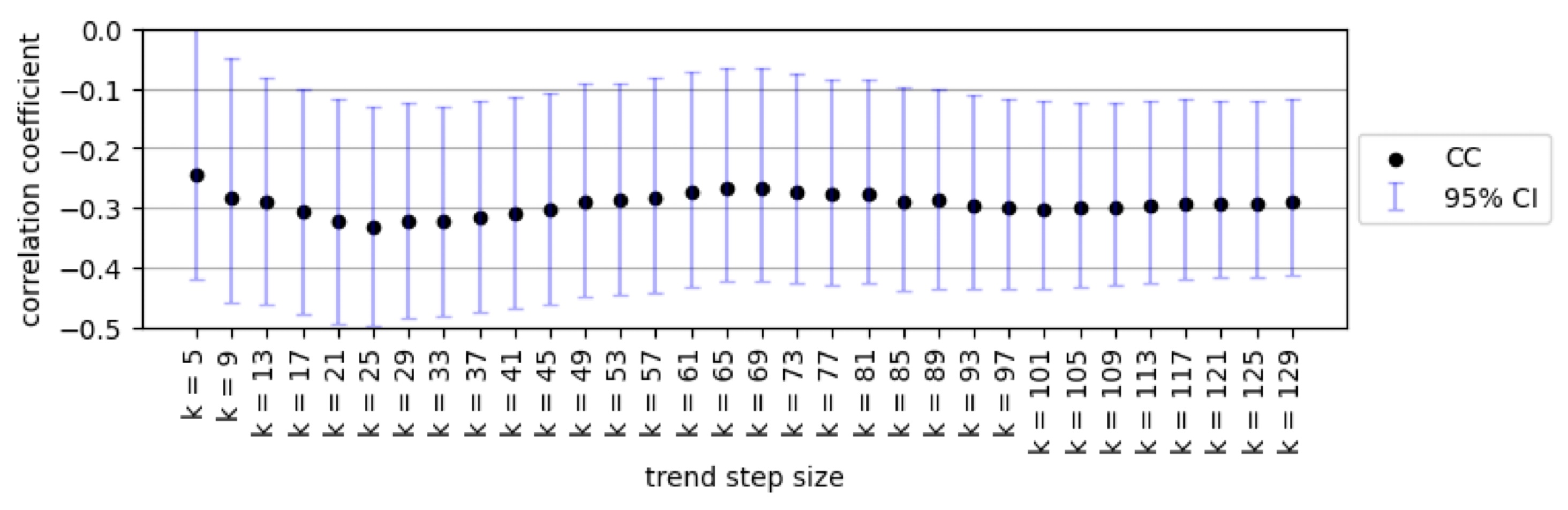}} \\
  (b) & \raisebox{-.5\height}{\includegraphics[width=.95\textwidth]{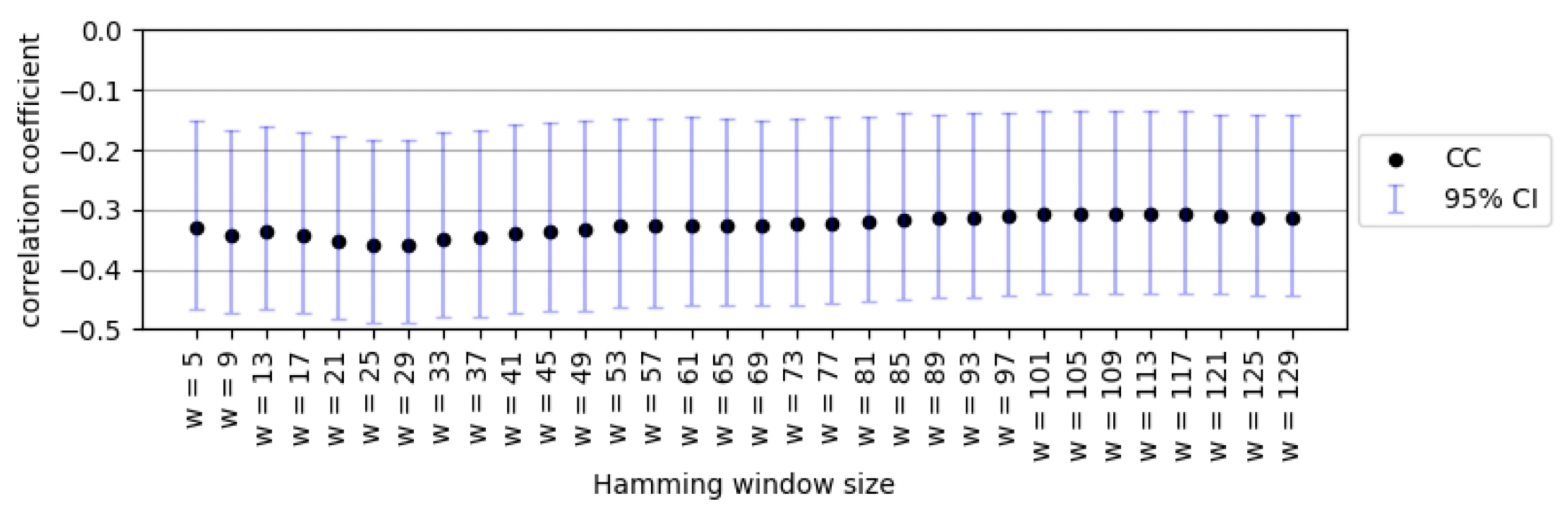}} \\
  (c) & \raisebox{-.5\height}{\includegraphics[width=.95\textwidth]{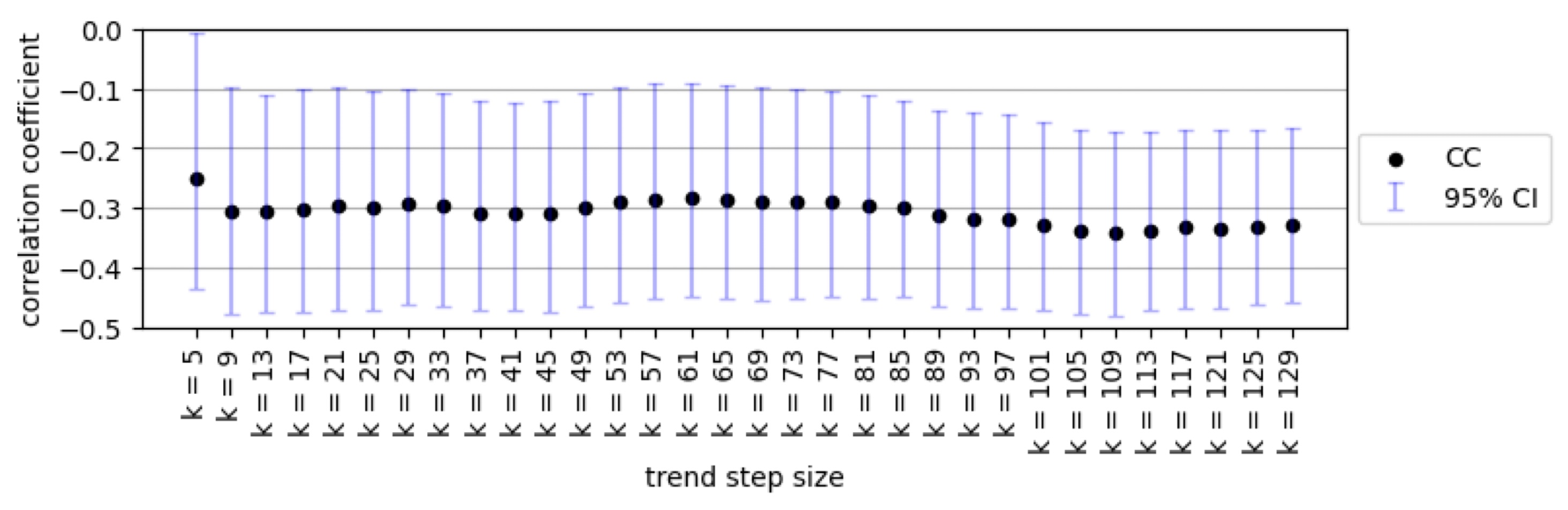}} \\
  (d) & \raisebox{-.5\height}{\includegraphics[width=.95\textwidth]{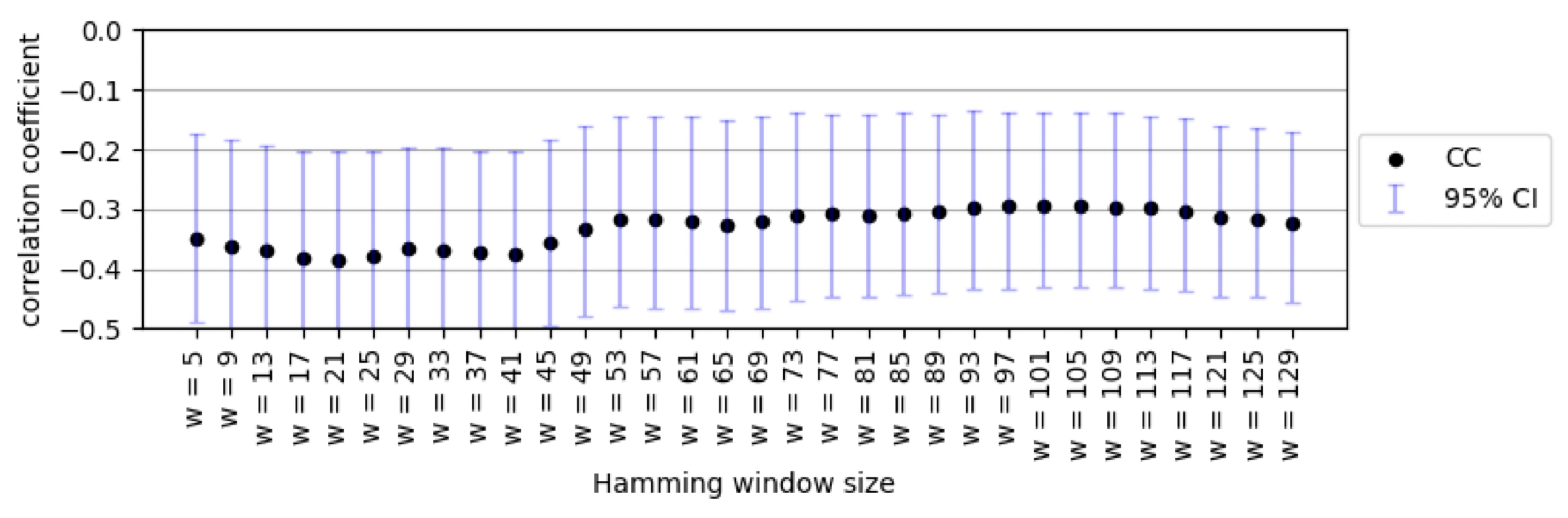}} \\
  \end{tabular}
  \caption{Figure (a) to (d) is the sensitivity analysis of model 1 to 4 respectively, presented by the Spearman correlation coefficients between various variability of morphology indices and MELD\_Na score.}
  \label{fig:sa}
\end{figure}

The sensitivity analysis of these four models is used to test the effect of different trend step size $k$ and Hamming window size $w$ to the four PPG models, and to determine whether the variability of morphology are sensitive to the behavior of the chosen parameter. The results of the sensitivity analysis by the CCs between various variability of morphology indices of the four models and the Meld\_Na score are shown in Figure \ref{fig:sa}. 
Note that for model 2 and 4, we first fix $w=0$ and test the parameter $k$, which have the same results of testing $k$ for model 1 and 3 respectively. Hence, we skip to show the sensitivity analysis of model 2 and 4 for testing the parameter $k$. Then, we fix $k$ that provide the best performance and test the parameter $w$ for each two models. 

The Spearman correlation between the four models of neohepatic phase and the Meld\_Na score are statistical significant among all $k$ from 5 to 129 (Figure \ref{fig:sa}(c)), expect of model 1 and model 2, which become significant from $k=9$ (Figure \ref{fig:sa}(a)). The variability of morphology of Model 1 and 3 reaches the best correlation at $k=25$ and $k=109$ respectively. 
As for the Spearman correlation between model 2 and 4 and the Meld\_Na score are statistical significant among all $w$ from 5 to 129 (Figure \ref{fig:sa}(b) and (d)). The variability of morphology of Model 2 and 4 reaches the best correlation at $w=29$ and $w=21$ respectively. 

As for confirming whether the variability of morphology show smooth curve for the four models, we use all the differences between adjacent CCs for quantifying the smoothness of the curve. 
The result of the differences will shown in the form $(a1, [a2, a3], a4)$, where $a1$ is the mean; $[a2, a3]$ is the minimum and maximum; and $a4$ is the IQR. 
For parameter $k$ of model 1 and 2, the differences between adjacent CCs for the correlations between MELD\_Na score and the variability of morphology are (0.006, [0.0002, 0.018], 0.005); For parameter $k$ of model 3 and 4, the differences are (0.006, [0.0004, 0.054], 0.003); And for parameter $w$ of model 2 and 4, the differences are (0.004, [0.0001, 0.014], 0.004) and (0.006, [0.0004, 0.021], 0.006) respectively. Note that we start from $k=9$ when calculating differences for parameter $k$ of model 1 and 2, since the CC of $k=5$ is not statistical significant.

The sensitivity analysis shows that all differences between adjacent CCs are small for the correlations between Meld\_Na and the variability of morphology of the four models, indicating the variability of morphology of all four models achieved consistent correlation and minimal fluctuation from a wide range of parameters $k$ and $w$. This result concludes that the derivations of the variability of morphology of all four models are robust and insensitive to the two input parameters. 
\section{Discussion}
Our results indicate that the variability of morphology obtained from PPG waveform correlates with clinical conditions at a level approaching that of ABP waveform. It suggests that the waveform signal data captured from a noninvasive sensor possess the information of the variability of morphology, which may grant more applications in the future.

\subsection{Signal processing perspective}
%The results support our methods. 
The Hamming window suppresses the fluctuation of the PPG waveform, while the employment of Wasserstein-1 distance alleviate the imperfection of the automatic pulse data isolation. Both provide improved metric for the affinity matrix in DDMap algorithm, which yields the manifold to quantify the variability of morphology. 

The DDMap algorithm used to calculate variability of morphology possesses theoretical robustness\cite{shen2022robust,shen2022scalability,lin2021wave}, which has been demonstrated in the sensitivity analysis also\cite{wang2023arterial}. Our methodology in this study not only improves performance but also maintains robustness as shown by sensitivity analysis, which enhances applicability in the future. 

It is worth mentioning that our PPG data benefited from several favorable conditions. The consistent anesthetic management throughout the surgery ensured the immobilization, maintained the adequate fluid status of the cardiovascular system, and stabilized the autonomic nerve system, all of which promote a favorable signal acquisition condition. It is important to caution that when employing a PPG sensor for various applications, these factors should be carefully considered. For example, violent physiological responses could be elicited by events such as stressful emotion, hungry or environmental temperature on a healthy human. Under such condition of peripheral vascular constriction, PPG is more susceptible than the direct ABP. 

\subsection{Biomedical perspective}

For clinical perspective, the results of PPG data show the variability of morphology in the neohepatic phase is associated with favorable clinical condition, which is in consistent with the ABP data counterpart in our previous study\cite{wang2023arterial}. As both PPG and ABP signal data are available during the surgery, timely assessment is an advantage over the laboratory examination. The beat-to-beat variation in waveform morphology presents both in both ABP waveform and PPG waveform imply other physical modalities of the cardiovascular waveform signal could capture the information, which is intrinsic in physiology. As the pulsatile waveform morphology is the summation of the wave traveling and reflecting throughout the vascular tree, we envision the signal data captured at different sites, whether upper limbs, lower limbs, cervical area, or their combination could provide more versatile application to reveal the human body condition.

The association between variability of morphology and clinical condition is reminiscent of HRV. While their notions seem similar at first glance, there are differences in physiology. The mechanism underlying HRV is mainly the cardiac sympathetic nerve system and the vagal nerve exerting opposite effect on the pacemaker of the heart -- the sinoatrial node. The variability of morphology could be regulated by the various controlling mechanism of the cardiovascular system, which could include the local blood flow regulation of several visceral organs, and the globalized (and possibly oversimplified) concepts of vascular tone and fluid status. The HRV is the variation of the instantaneous heart rate, an one-dimensional time series, while the successive pulse waveform is multivariate time series, which mandates our methodology\cite{wang2023arterial}. Therefore seeing them as the extensions of the heart rate and blood pressure, we speculate HRV has more direct association to autonomic nerve system while the waveform variability of morphology is more related with the cardiovascular system.

% Also, note that the postsurgical scores L-GrAFT10 and EASE are evaluated using the laboratory data within three month after the operation, which are not predictions. However, the correlations between variability of morphology of the neohepatic phase, which can be calculated immediately after the surgery, and L-GrAFT10 and EASE scores gives us an insight into the patient's prognostic condition following the transplantation, which may lead us to a better postoperative care for those who has low variability of morphology and probably a worse recovery status.

\subsection{Limitation and applicability}

Although our PPG results are encouraging, extending the variability of morphology toward more applications may require some consideration of its limitations. It may require stable physiological conditions and minimal interference of sensor data acquisition, requiring the finger or hand to be as still as possible.

Despite the differences, there should be intangible physiological interaction between HRV and the variability of morphology in cardiovascular waveform. Therefore, we anticipate that the combination of the two may provide a more comprehensive assessment in humans, worthy of future studies.
\section{Conclusion}

The beat-to-beat variability of morphology in PPG signal data presents the association with clinical condition via quantification based on unsupervised manifold learning. 

\section{Acknowledgement}
We thank Professor Hau-Tieng Wu (Department of Mathematics, New York University, New York) for the valuable suggestion on the signal artifact problems that greatly assisted the research. The work is supported by the funding of National Science and Technology Council, Taipei, Taiwan (112-2115-M-075 -001 - , 112-2314-B-075 -070 -MY2 ).

%% ---------------------------
%% Bibliography
%% ---------------------------

\bibliographystyle{IEEEtran}
\bibliography{article}

\begin{thebibliography}{10}
\providecommand{\url}[1]{#1}
\csname url@samestyle\endcsname
\providecommand{\newblock}{\relax}
\providecommand{\bibinfo}[2]{#2}
\providecommand{\BIBentrySTDinterwordspacing}{\spaceskip=0pt\relax}
\providecommand{\BIBentryALTinterwordstretchfactor}{4}
\providecommand{\BIBentryALTinterwordspacing}{\spaceskip=\fontdimen2\font plus
\BIBentryALTinterwordstretchfactor\fontdimen3\font minus
  \fontdimen4\font\relax}
\providecommand{\BIBforeignlanguage}[2]{{%
\expandafter\ifx\csname l@#1\endcsname\relax
\typeout{** WARNING: IEEEtran.bst: No hyphenation pattern has been}%
\typeout{** loaded for the language `#1'. Using the pattern for}%
\typeout{** the default language instead.}%
\else
\language=\csname l@#1\endcsname
\fi
#2}}
\providecommand{\BIBdecl}{\relax}
\BIBdecl

\bibitem{teboul2016less}
J.-L. Teboul, B.~Saugel, M.~Cecconi, D.~De~Backer, C.~K. Hofer, X.~Monnet,
  A.~Perel, M.~R. Pinsky, D.~A. Reuter, A.~Rhodes \emph{et~al.}, ``Less
  invasive hemodynamic monitoring in critically ill patients,'' \emph{Intensive
  care medicine}, vol.~42, pp. 1350--1359, 2016.

\bibitem{avolio2009role}
A.~P. Avolio, L.~M. Van~Bortel, P.~Boutouyrie, J.~R. Cockcroft, C.~M. McEniery,
  A.~D. Protogerou, M.~J. Roman, M.~E. Safar, P.~Segers, and H.~Smulyan, ``Role
  of pulse pressure amplification in arterial hypertension: experts’ opinion
  and review of the data,'' \emph{Hypertension}, vol.~54, no.~2, pp. 375--383,
  2009.

\bibitem{hatib2018machine}
F.~Hatib, Z.~Jian, S.~Buddi, C.~Lee, J.~Settels, K.~Sibert, J.~Rinehart, and
  M.~Cannesson, ``Machine-learning algorithm to predict hypotension based on
  high-fidelity arterial pressure waveform analysis,'' \emph{Anesthesiology},
  vol. 129, no.~4, pp. 663--674, 2018.

\bibitem{wang2023arterial}
S.-C. Wang, C.-K. Ting, C.-Y. Chen, C.~Liu, N.-C. Lin, C.-C. Loong, H.-T. Wu,
  and Y.-T. Lin, ``Arterial blood pressure waveform in liver transplant surgery
  possesses variability of morphology reflecting recipients’ acuity and
  predicting short term outcomes,'' \emph{Journal of Clinical Monitoring and
  Computing}, pp. 1--11, 2023.

\bibitem{lin2021intraoperative}
Y.-T. Lin, H.-T. Wu, S.-C. Wang, C.-K. Ting, C.~Liu, N.-C. Lin, C.-Y. Chen, and
  C.-C. Loong, ``Intraoperative arterial pressure waveforms shows temporal
  structure complexity correlated with acuity of liver transplant by pulse wave
  manifold learning analysis,'' in \emph{Anesthesia \& Analgesia}, vol. 132,
  2021, pp. 15--16.

\bibitem{wang2020novel}
S.-C. Wang, H.-T. Wu, P.-H. Huang, C.-H. Chang, C.-K. Ting, and Y.-T. Lin,
  ``Novel imaging revealing inner dynamics for cardiovascular waveform analysis
  via unsupervised manifold learning,'' \emph{Anesthesia \& Analgesia}, vol.
  130, no.~5, pp. 1244--1254, 2020.

\bibitem{shen2022robust}
C.~Shen, Y.-T. Lin, and H.-T. Wu, ``Robust and scalable manifold learning via
  landmark diffusion for long-term medical signal processing,'' \emph{J. Mach.
  Learn. Res.}, vol.~23, pp. 1--30, 2022.

\bibitem{vlachopoulos2011mcdonald}
C.~Vlachopoulos, M.~O'Rourke, and W.~W. Nichols, \emph{McDonald's blood flow in
  arteries: theoretical, experimental and clinical principles}.\hskip 1em plus
  0.5em minus 0.4em\relax CRC press, 2011.

\bibitem{lin2021wave}
Y.-T. Lin, J.~Malik, and H.-T. Wu, ``Wave-shape oscillatory model for
  nonstationary periodic time series analysis,'' \emph{Foundations of Data
  Science}, vol.~3, no.~2, pp. 99--131, 2021.

\bibitem{sluyter2019identification}
J.~D. Sluyter, A.~D. Hughes, C.~A. Camargo~Jr, S.~A.~M. Thom, K.~H. Parker,
  B.~Hametner, S.~Wassertheurer, and R.~Scragg, ``Identification of distinct
  arterial waveform clusters and a longitudinal evaluation of their clinical
  usefulness,'' \emph{Hypertension}, vol.~74, no.~4, pp. 921--928, 2019.

\bibitem{lin2019unexpected}
Y.-T. Lin, Y.-L. Lo, C.-Y. Lin, M.~G. Frasch, and H.-T. Wu, ``Unexpected
  sawtooth artifact in beat-to-beat pulse transit time measured from patient
  monitor data,'' \emph{PloS one}, vol.~14, no.~9, p. e0221319, 2019.

\bibitem{shen2022scalability}
C.~Shen and H.-T. Wu, ``Scalability and robustness of spectral embedding:
  landmark diffusion is all you need,'' \emph{Information and Inference: A
  Journal of the IMA}, vol.~11, no.~4, pp. 1527--1595, 2022.

\bibitem{zelnik2004self}
L.~Zelnik-Manor and P.~Perona, ``Self-tuning spectral clustering,''
  \emph{Advances in neural information processing systems}, vol.~17, 2004.

\bibitem{vaserstein1969markov}
L.~N. Vaserstein, ``Markov processes over denumerable products of spaces,
  describing large systems of automata,'' \emph{Problemy Peredachi
  Informatsii}, vol.~5, no.~3, pp. 64--72, 1969.

\bibitem{smith2011spectral}
J.~O. Smith~III, \emph{Spectral audio signal processing}.\hskip 1em plus 0.5em
  minus 0.4em\relax W3K publishing, 2011.

\bibitem{brown2002model}
R.~S. Brown~Jr, K.~S. Kumar, M.~W. Russo, M.~Kinkhabwala, D.~L. Rudow,
  P.~Harren, S.~Lobritto, and J.~C. Emond, ``Model for end-stage liver disease
  and child-turcotte-pugh score as predictors of pretransplantation disease
  severity, posttransplantation outcome, and resource utilization in united
  network for organ sharing status 2a patients,'' \emph{Liver Transplantation},
  vol.~8, no.~3, pp. 278--284, 2002.

\bibitem{biggins2006evidence}
S.~W. Biggins, W.~R. Kim, N.~A. Terrault, S.~Saab, V.~Balan, T.~Schiano,
  J.~Benson, T.~Therneau, W.~Kremers, R.~Wiesner \emph{et~al.},
  ``Evidence-based incorporation of serum sodium concentration into meld,''
  \emph{Gastroenterology}, vol. 130, no.~6, pp. 1652--1660, 2006.

\bibitem{agopian2021multicenter}
V.~G. Agopian, D.~Markovic, G.~B. Klintmalm, G.~Saracino, W.~C. Chapman,
  N.~Vachharajani, S.~S. Florman, P.~Tabrizian, B.~Haydel, D.~Nasralla
  \emph{et~al.}, ``Multicenter validation of the liver graft assessment
  following transplantation (l-graft) score for assessment of early allograft
  dysfunction,'' \emph{Journal of hepatology}, vol.~74, no.~4, pp. 881--892,
  2021.

\bibitem{agopian2018evaluation}
V.~G. Agopian, M.~P. Harlander-Locke, D.~Markovic, W.~Dumronggittigule, V.~Xia,
  F.~M. Kaldas, A.~Zarrinpar, H.~Yersiz, D.~G. Farmer, J.~R. Hiatt
  \emph{et~al.}, ``Evaluation of early allograft function using the liver graft
  assessment following transplantation risk score model,'' \emph{JAMA surgery},
  vol. 153, no.~5, pp. 436--444, 2018.

\bibitem{avolio2020development}
A.~W. Avolio, A.~Franco, A.~Schlegel, Q.~Lai, S.~Meli, P.~Burra, D.~Patrono,
  M.~Ravaioli, D.~Bassi, F.~Ferla \emph{et~al.}, ``Development and validation
  of a comprehensive model to estimate early allograft failure among patients
  requiring early liver retransplant,'' \emph{JAMA surgery}, vol. 155, no.~12,
  pp. e204\,095--e204\,095, 2020.

\end{thebibliography}

\end{document}